\documentclass[%
showkeys,
 reprint,
 amsmath,amssymb,
 aps,
 floatfix,
]{revtex4-2}

\usepackage{graphicx}
\usepackage{dcolumn}
\usepackage{bm}
\usepackage[table]{xcolor}
\usepackage{colortbl}  
\usepackage{pgffor} 

\begin{document}

\preprint{APS/123-QED}

\title{Elasticity-mediated Morphogenesis in Interfacial Colloidal Assemblies}

\author{Vaibhav Raj Singh Parmar}
\author{Sayantan Chanda}%

\author{Rituparno Mandal}
 
\author{Ranjini Bandyopadhyay}
\affiliation{Soft Condensed Matter Group, Raman Research Institute, C. V. Raman Avenue, Sadashivanagar, Bangalore 560 080, INDIA}%

\date{\today}

\begin{abstract}
We study the self-assembly of colloidal microgel particles at a quasi-two-dimensional air-water interface of a drying droplet. Using bright-field microscopy, we demonstrate that increasing particle elasticity drives interfacial organization from repulsion-stabilized crystallization to attraction-dominated gelation, via diverse metastable structures including clusters, voids and anisotropic aggregates. Molecular dynamics simulations using an effective potential that captures the interplay between hydrophobic, capillary, steric and dipolar interactions, reproduce the overall phenomenology of the observed colloidal morphogenesis. Our findings establish particle elasticity as a key parameter governing non-equilibrium structural organization of colloids at an interface.
\end{abstract}
\maketitle

Non-equilibrium structural organization, a key feature of biological and soft-matter systems, drives phase transitions and the formation of complex morphologies across different spatiotemporal scales~\cite{doi:10.1126/science.1070821}. 
While rigid colloids have provided significant insights into 2D phase transitions and interfacial self-assembly~\cite{JMKosterlitz_1973,10.1063/1.3188948,PhysRevLett.85.3656}, they are insufficient to capture the structural evolution exhibited by deformable colloids, where internal degrees of freedom and particle elasticity can drastically modulate inter-particle interactions~\cite{PhysRevX.10.031012, Bergman2018, doi:10.1021/acsnano.1c02486, doi:10.1021/acs.accounts.9b00528}.
Due to their core-corona architecture~\cite{PhysRevLett.131.258202}, polymer-based soft colloidal microgels are excellent model systems for exploring how particle deformability influences non-equilibrium structural organization in glassy systems~\cite{Bergman2018}, non-Newtonian flows~\cite{10.1063/5.0232833} and crystallization~\cite{doi:10.1021/jacs.7b08503,PhysRevLett.114.098303}.
 
 \vspace{-0.05cm}
 Microgels spontaneously adsorb at fluid interfaces due to their amphiphilic nature~\cite{https://doi.org/10.1002/admi.201500371} while also deforming to reduce interfacial tension~\cite{JOSE2023364,https://doi.org/10.1002/admi.201500371,doi:10.1021/acs.accounts.9b00528}. During the drying of sessile droplets, these adsorbed microgels inhibit the coffee-ring effect~\cite{Deegan1997} and form various 2D structures dictated by local physicochemical conditions~\cite{https://doi.org/10.1002/admi.201500371, DESHMUKH2015215}.
Beyond its fundamental role in probing phase transitions and the organization of particles into higher order complex structures~\cite{10.1063/1.3188948, Vaibhav2025,Naz2024,JOSE2023364,MAYARANI2021683,D4NA00542B}, interface-assisted self-assembly of colloidal particles offers a robust platform for stabilizing foams and emulsions~\cite{doi:10.1021/la203062b,doi:10.1021/la302331s} and in nanolithography~\cite{doi:10.1021/acsnano.1c02486}. While the role of elasticity in dried microgel  deposits has been well-documented~\cite{D1SM00841B,https://doi.org/10.1002/admi.201500371}, its influence on interparticle interactions and dynamic self-assembled architectures remains largely unexplored. The air-water interface of evaporating sessile droplets can potentially provide a simple way to capture such elasticity mediated change in effective interaction and its role in colloidal morphogenesis.


 \vspace{-0.05cm}
In this Letter, we demonstrate that changes in particle elasticity drive a series of distinct morphological transitions in the non-equilibrium structural organization of poly(N-isopropylacrylamide) (PNIPAM) microgels~\cite{PELTON1986247}. By drying sessile droplets of dilute suspensions, we harnessed spatial gradients in microgel concentration at the droplet interface to study self-assembly over a continuous range of local area fractions, $\phi$. We report that soft, highly deformable microgels transform from isolated circular clusters to foam-like void structures, eventually stabilizing into ordered, faceted hexagonal lattices at high $\phi$. In contrast, increasing microgel crosslinker density enhances particle elasticity, promoting anisotropic structures such as chain-like aggregates that percolate into 2D disordered gel-like networks. We quantify the observed morphological diversity through the global bond orientational order parameter, $|\psi_6|$, and pair correlation function, $g(r)$. We propose an effective interparticle potential, $U(r)$, incorporating hydrophobic, steric, capillary and dipolar electrostatic contributions~\cite{Fernandez2006,polym13091353,Cohin2013,PhysRevLett.45.569}. 
Tuning the steric range and dipolar strength in molecular dynamics (MD) simulations allows us to replicate the entire array of experimental morphologies. 
Our results reveal that microgel elasticity fundamentally alters the interparticle potential, providing a tunable pathway for designing complex 2D architectures.

\begin{figure}[!t]
	\includegraphics[width= 3.4in]{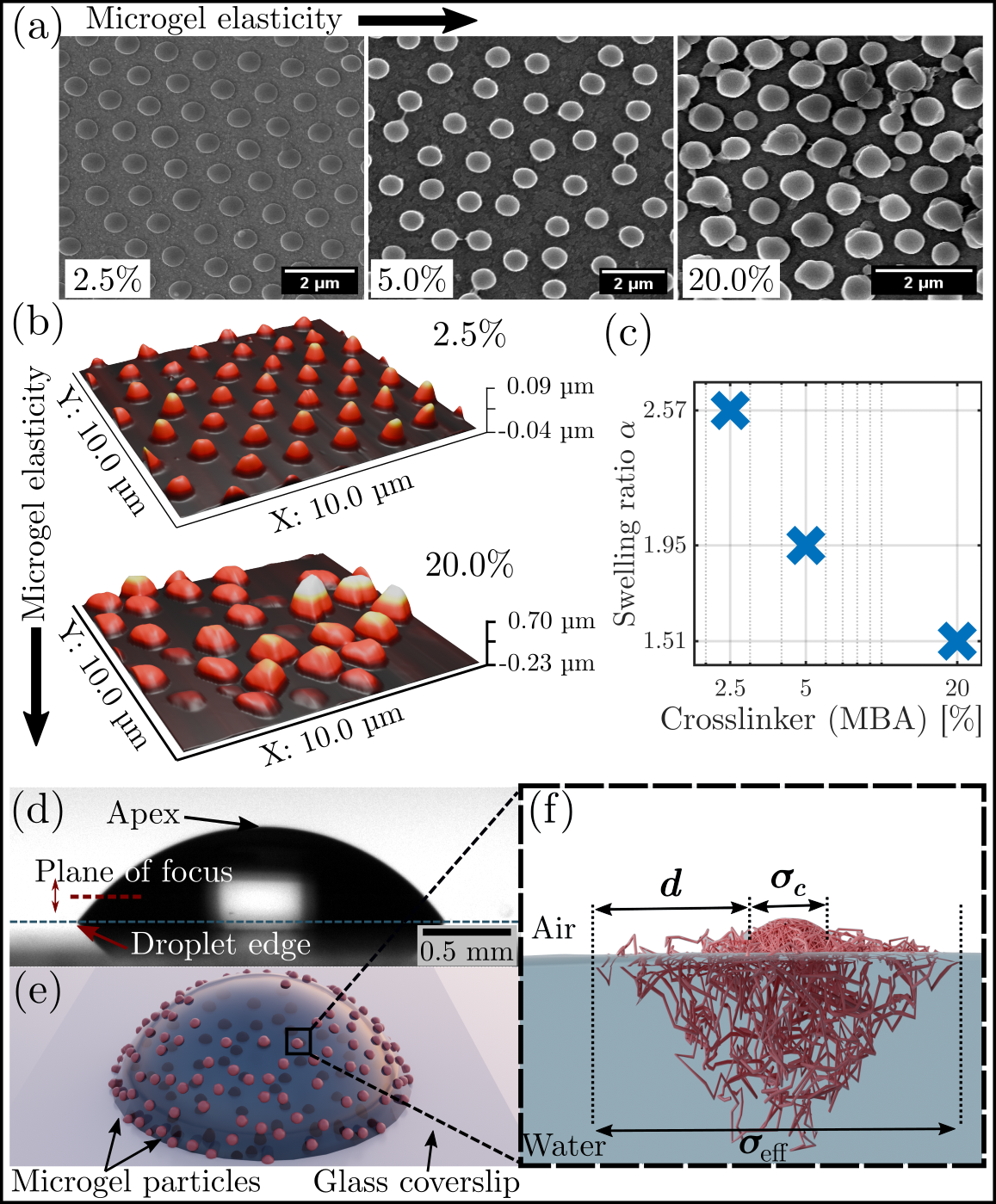}
		\centering
        \caption{\label{fig:1} (a) SEM images of dried microgels at different crosslinker (MBA) densities. (b) AFM height profiles of dried microgels of two different crosslinker densities. (c) $\alpha$ vs crosslinker density. (d) Side-view micrograph of a sessile microgel droplet on a glass substrate. (e) Schematic illustration of the droplet and adsorbed microgel particles (purple spheres). (f) Zoomed-in view of a single adsorbed microgel at the air-water interface. The effective diameter $\sigma_{\text{eff}}$ of a microgel is labeled as the sum of the diameter of the highly crosslinked core ($\sigma_{c}$) and twice the thickness ($d$) of the deformable corona.}
\end{figure}

 \vspace{-0.05cm}
PNIPAM microgels of different crosslinker densities were synthesized (Supplementary Section~ST1)~\cite{PELTON1986247}. Scanning electron microscopy (SEM) images in Fig.~\ref{fig:1}(a) display microgels synthesized over a range of crosslinker densities. Height profiles in Fig.~\ref{fig:1}(b) show that a lower crosslinker density leads to reduced heights of the dried microgels, indicating higher deformability~\cite{D1SM00841B}. Fig.~\ref{fig:1}(c) shows that the swelling ratio $\alpha$~\cite{10.1063/5.0232833}, the ratio of mean hydrodynamic diameters of the microgels in the maximally swollen ($20^\circ\text{C}$) and collapsed ($45^\circ\text{C}$) states (Supplementary Section~ST2), decreases with increasing crosslinker (MBA) content. This reduction in $\alpha$ implies a higher elastic modulus for stiffer microgels.
We dried 1$\mu$L sessile droplets of microgel suspensions (particle concentrations $\sim 0.002$--$0.008$\%w/v
) on glass coverslips under ambient conditions (temperature = 22 $\pm$ 2$^\circ$C and relative humidity $\approx$ 40-45\%). A side-view micrograph of a droplet, imaged using a contact angle goniometer, is shown in Fig.~\ref{fig:1}(d) and a schematic is shown in Fig.~\ref{fig:1}(e). Owing to their core-corona architecture, the microgels flatten into a fried egg configuration (Fig.~\ref{fig:1}(f)) to minimize interfacial energy. The diameter of the incompressible core and thickness of the compressible corona are labeled as ($\sigma_{c}$) and ($d$) respectively in Fig.~\ref{fig:1}(f). Microgel self-assembly at the air-water interface during droplet evaporation (Supplementary Fig.~S3) was monitored in bright-field transmission mode with a 63X oil-immersion objective. These images were further processed using Ilastik~\cite{berg2019} and MATLAB (Supplementary Fig.~S4).
 \vspace{-0.04cm}
Microgels in evaporating sessile droplets populate the air-water interface via two synergistic routes. First, capillary flows transport the colloids toward the droplet edge (triple phase contact line, TPCL) ~\cite{Deegan1997}, establishing a spatial gradient in $\phi$ that decreases from the TPCL toward the apex. 
Second,  $\phi$ continues to increase along the entire interface due to evaporation-induced water loss and migration of microgels from the bulk. In our experiments, the spatial variation of $\phi$ was significantly more pronounced than its temporal increase. 
This experimental design allowed us to investigate a wide range of distinct assembly processes at the interface of the evaporating droplet. Diverse metastable self-assembled structures were observed to coexist and evolve temporally due to microgel rearrangements and enhancement of $\phi$ owing to evaporation. Representative reconstructed micrographs of self-assembled microgels of different elasticities are displayed in Fig~\ref{fig:2}, with corresponding raw images provided in Supplementary Fig.~S5. The microgels are color-coded by bond orientational order $|\psi_6|$ to highlight the local positional order of the assemblies. The calculated distributions of $|\psi_6|$ are presented in Supplementary Fig.~S6. 

 \vspace{-0.05cm}
\begin{figure}[t]
	\includegraphics[width= 3.4in]{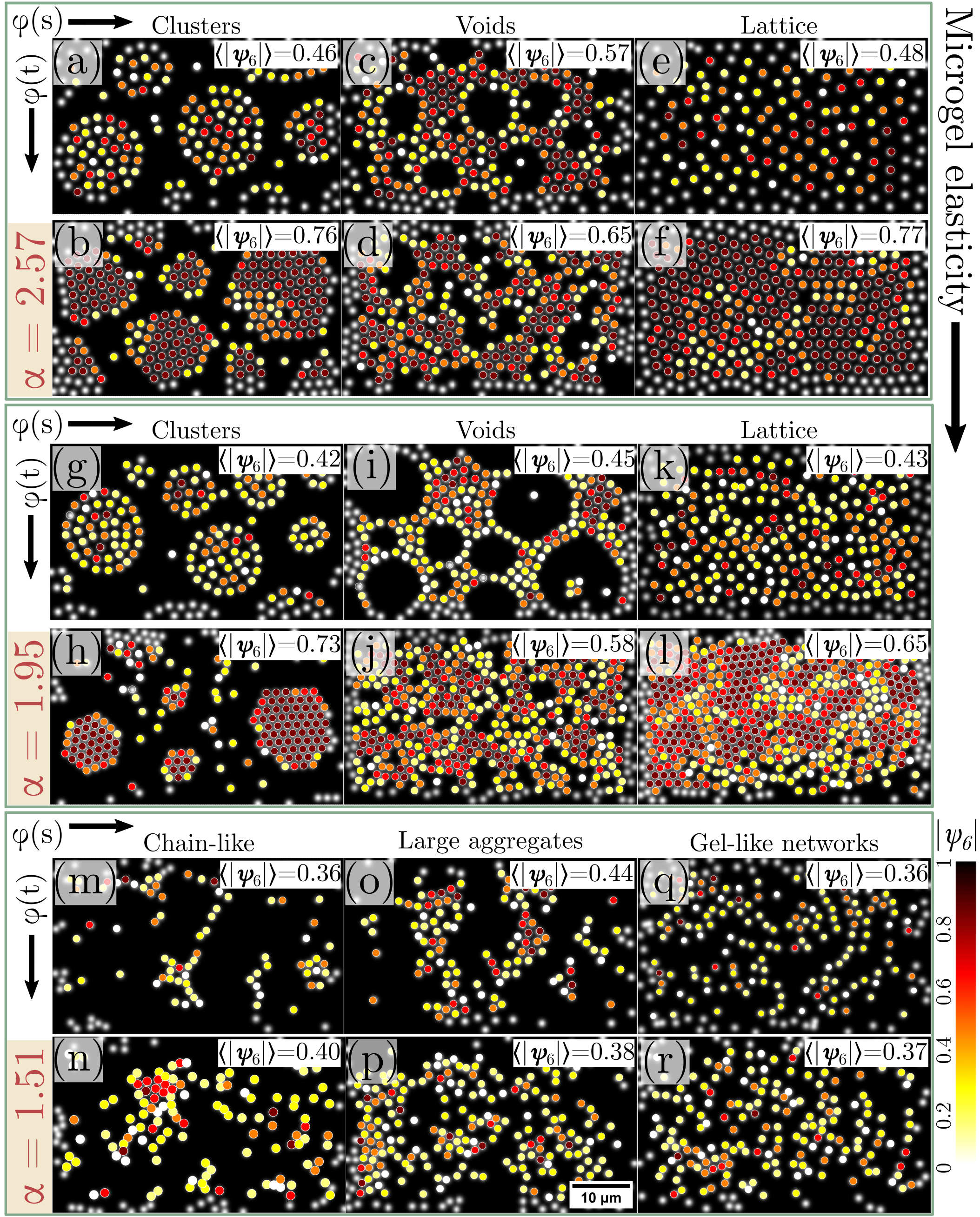}
	\centering
	\caption{\label{fig:2} Reconstructed bright-field micrographs showing the spatiotemporal self-assembly of microgel cores for three particle elasticity values (top to bottom panels). Top panel (a-f): soft microgels ($\alpha$ = 2.57). Middle panel (g-l): microgels of intermediate elasticity ($\alpha$ = 1.95). Bottom panel (m-r): stiff microgels ($\alpha$ = 1.51). The local area fraction ($\phi$, denoted by $\phi$(s)) increases from left to right as the triple phase contact line of the droplet is approached. Within each panel, $\phi$ also increases from top to bottom (earlier and later time data, denoted by $\phi$(t)) due to evaporative water loss. Microgels are color-coded by the local hexagonal bond orientational order parameter, $|\psi_6|$.  Scale bar at bottom right is 10 $\mu$m.}
\end{figure}
Soft microgels ($\alpha$ = 2.57) characterized by low elasticity (Fig.~\ref{fig:1}(c)) assembled into circularly ordered colloidal clusters  (average bond orientational order parameter $\langle |\psi_6| \rangle$ = 0.46) at low $\phi$ (Fig.~\ref{fig:2}(a)). During the course of droplet evaporation, these circular clusters transitioned to faceted structures ($\langle |\psi_6| \rangle = 0.76$), displayed in Fig.~\ref{fig:2}(b). Time-lapse images of cluster coalescence are displayed in Supplementary Fig.~S7. With increase in $\phi$ as the TPCL was approached, 2D foam-like structures with circular voids surrounded by microgel walls were observed (Fig.~\ref{fig:2}(c)). Interestingly, circular voids coalesced to form bigger voids (Supplementary Fig.~S8) and transformed to non-circular shapes as the microgels ordered with elapsed time, evident from the gradually increasing value of $\langle |\psi_6| \rangle$ in Figs.~\ref{fig:2}(c,d). At the highest $\phi$, the soft microgel particles self-assembled into loosely arranged hexagonal lattices (Figs.~\ref{fig:2}(e,f)), with $\langle |\psi_6| \rangle \approx 0.77$. The gradual increase in $\langle |\psi_6| \rangle$ with $\phi$ indicates a systematic transition to more ordered assemblies. 
 At early times, microgels of intermediate elasticity  ($\alpha=1.95$) exhibited self-assembled structures (Figs.~\ref{fig:2}(g-l)) that were qualitatively similar to their softer counterparts ($\alpha=2.57$). At higher $\phi$, however, the structures showed clear deviations, with the observed hexagonal structures characterized by a lower degree of order. Exotic configurations, such as microgel clusters coexisting with chain-like aggregates and highly ordered domains coexisting with disordered assemblies and 2D foams, were also observed in this intermediate regime, as shown in Supplementary Fig.~S9. 
Highly elastic microgel particles ($\alpha=1.51$) exhibited markedly different self-assembled morphologies. Instead of forming ordered clusters, these particles assembled into chain-like aggregates at low $\phi$ (Fig.~\ref{fig:2}(m)). As shown in Supplementary Fig.~S10, the self-assembled morphologies evolved by folding and interconnecting with neighboring aggregates (Fig.~\ref{fig:2}(n)). With increasing $\phi$, the chains transformed into anisotropic clusters (Fig.~\ref{fig:2}(o)) that eventually percolated to form disordered 2D gel-like networks (Figs.~\ref{fig:2}(p-q)). These structures are reminiscent of those seen in simulations of strongly attractive colloids~\cite{doi:10.1126/sciadv.abb8107}.

\vspace{-0.08cm}
Several studies have reported the emergence of complex colloidal mesostructures, such as foams, cellular structures, ordered aggregates and voids, when colloidal particles were spread at air–water or oil-water interfaces~\cite{PhysRevE.58.660,Nikolaides2002,doi:10.1021/la0497090,PhysRevE.62.5263}. Contact line undulations, arising from surface roughness or chemical inhomogeneity of the adsorbed microgels, have been shown to result in interparticle capillary attractions~\cite{PhysRevE.62.5263}. In addition, an effective steric repulsion arises due to compression of the polymer brushes comprising microgel coronas~\cite{PhysRevX.10.031012}. For the morphologies formed by soft microgels, the interparticle separation exceeds the particle hydrodynamic diameter (Figs.~\ref{fig:2}(a-f), Supplementary Fig.~S11), suggesting that the structures are stabilized by steric repulsion. For stiff microgel assemblies, the average interparticle separation is comparable to the particle diameter (Figs.~\ref{fig:2}(m-r), Supplementary Fig.~S11), suggesting that short-ranged attractions control self-assembly. Microgels consist of polymers comprising both hydrophilic and hydrophobic moieties. These two parts interact differently with the surrounding aqueous medium, thereby changing the structure of the water molecules near the particle surface. Consequently, a purely entropic short-ranged hydrophobic attraction arises between neighboring microgel cores. A repulsive shoulder in the interparticle potential prevents particle close packing and stabilizes anisotropic structures~\cite{PhysRevE.58.1478,doi:10.1021/jacs.7b08503}. Microgels at fluid interfaces behave as dipoles due to the uneven screening of surface charges at the interface. This can give rise to a long-range repulsive shoulder in the interparticle potential. In our experiments with stiff colloids, a weak steric barrier, arising from small $d$ values, promotes the formation of anisotropic aggregates and percolated gel-like networks. Stiff microgels primarily interact via hydrophobic attraction, while a long-range dipolar repulsion ensures their stability.
\begin{table}[b!]
  \centering
  \caption{Simulation parameters employed for different microgel elasticity values. The parameters corona thickness, $d$, and dipolar repulsion strength, $p$, were varied in the simulations while keeping all other parameters fixed to mimic different scenarios with $\alpha$ = 2.57, 1.95 and 1.51.}
  \label{tab:1}
  \begin{tabular}{|c|c|c|c|}
    \hline
      & Corona thickness & Dipolar strength& Swelling ratio\\
     \hline
      & $d$ & $p$&$\alpha$\\
    \hline
    Case I  & 0.35 & 0.2& 2.57\\
    Case II  & 0.22 & 3.0& 1.95\\
    Case III & 0.05 & 6.0& 1.51\\
    \hline
  \end{tabular}
\end{table}

 \vspace{-0.05cm}
To elucidate the interplay of competing interactions driving these self-assembled morphologies, we propose an effective pair potential, $U(r)$, as a function of center-to-center microgel separation, $r$:
\begin{equation}
\label{eq:1}
    U(r) =
\begin{cases}
\infty,  & r \leq \sigma_{\text{c}}, \\
 U_{h} + U_{s} + U_{c} + U_{d}, & \sigma_{\text{c}} < r < 8\sigma_{\text{c}}.\\
 0, & r > 8\sigma_{\text{c}}
\end{cases}
\end{equation}
In the above expression, the hard wall repulsion at $r \leq \sigma_{\text{c}}$ arises from volume exclusion of the microgel cores. The different contributions to $U(r)$ for $\sigma_{\text{c}} < r < 8\sigma_{\text{c}}$ are:\\
(i) {\textit{Hydrophobic attraction}:} This is modeled as $U_{h} = - W_0 \exp{-(h/\lambda_0)}$~\cite{Fernandez2006}, where $W_0$ and $\lambda_0$ denote the strength and range of the interaction, and $h = r - 2\sigma_c$ is the surface separation between two microgel cores. \\
(ii) {\textit{Steric repulsion}:} This can be described by a two-dimensional Hertzian expression: $U_{s} = S\sigma_{\text{eff}}^2\left(1 - \frac{r}{\sigma_{\text{eff}}} \right)^2 \Theta(\sigma_{\text{eff}} - r)$~\cite{PhysRevX.10.031012}, where $S$ is the steric repulsion strength, $\sigma_{\text{eff}} = \sigma_{\text{c}} + 2d$ is the effective particle diameter (Fig.~\ref{fig:1}(f)) and $\Theta$ is the Heaviside function.\\
(iii) {\textit{Capillary attraction}:}  We use $U_{c} = - \frac{B}{r^4 + 1}$~\cite{DANOV2005121} to model quadrupolar capillary attraction. Here, $B$ represents attraction strength, and unity was added in the denominator ad hoc to prevent divergence of the capillary attraction at small $r$.\\
(iv) {\textit{Electrostatic repulsion}:}  $U_{d} = p/r^3$~\cite{PhysRevLett.45.569} describes the dipolar repulsion between two microgel particles. Here, $p$ represents the strength of the long-range dipolar repulsion.

 \vspace{-0.04cm}
To corroborate the effective potential in Eq.~\ref{eq:1}, we performed 2D MD simulations in LAMMPS~\cite{LAMMPS}, using an NVT ensemble and a Langevin thermostat (Supplementary Section~ST3). 
The following parameters were maintained at fixed values: $\sigma_{\text{c}} = 1.0$,  $W_0 = 8.0$, $\lambda_0 = 0.01$, $B = 10.0$, and $S = 19.0$). Consequently, only $d$ and $p$ were varied to model microgels of varying elasticities (Table~\ref{tab:1}). The choice of parameters was based on the underlying physical principles, while ensuring that the observed structures were replicated. We note that this parameter space is not unique, and other combinations exist that can reproduce the experimentally observed assemblies.  Since all the experimentally observed self-assembled morphologies could be modeled by varying the corona thickness and dipolar strength, additional contributions of drying-induced flows such as capillary and Marangoni flows were not considered in our simulations.
\begin{figure}[t]
	\includegraphics[width= 3.25in]{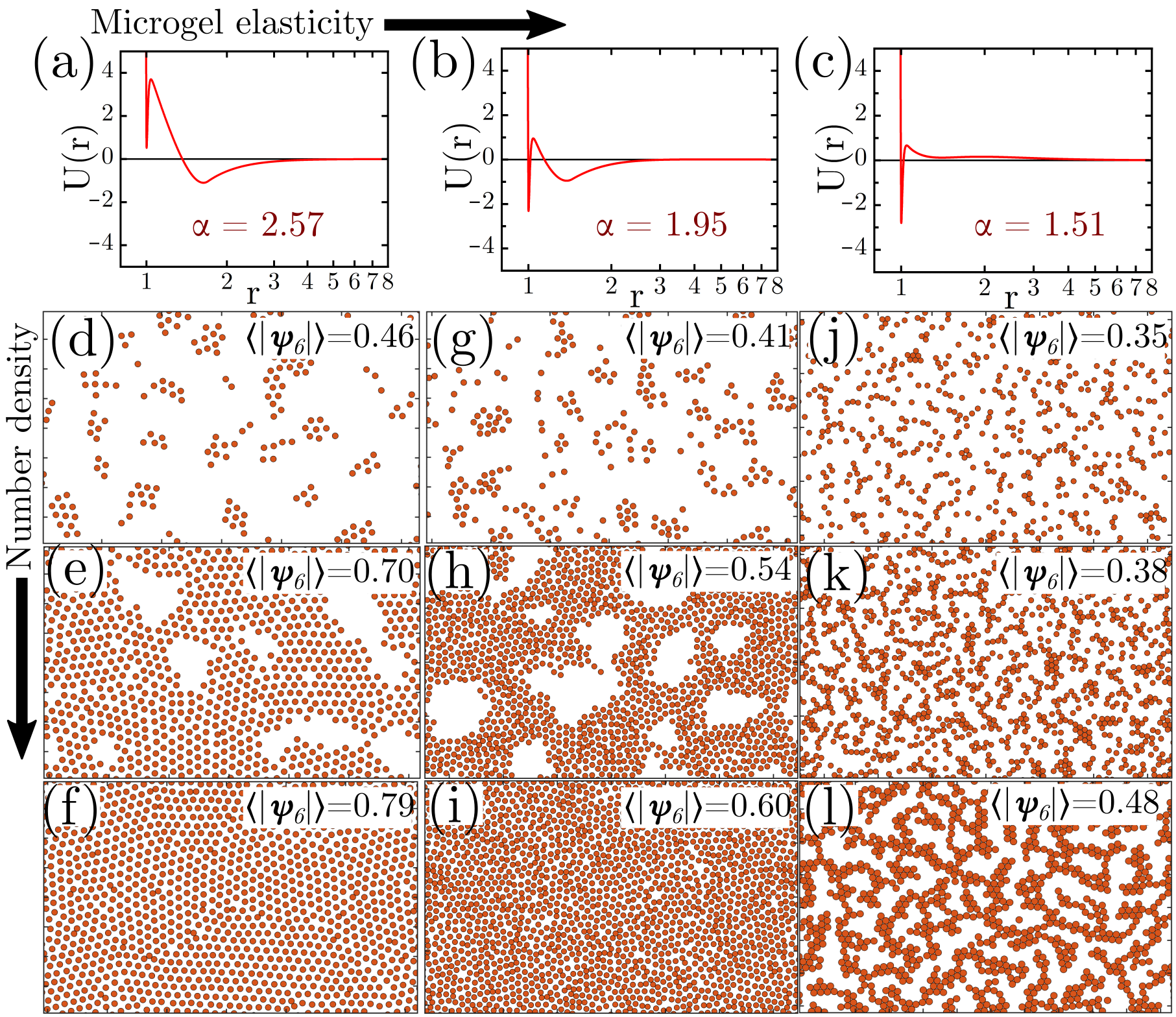}
	\centering
	\caption{\label{fig:3} Effective pair potentials, $U(r)$ (in units of $k_BT$), as a function of interparticle separation, $r$, for (a) $\alpha$ = 2.57; soft microgels, (b) $\alpha$ = 1.95; microgels of intermediate elasticity and (c) $\alpha$ = 1.51; stiff microgels. Snapshots obtained from MD simulations showing distinct microgel assemblies with increasing particle number densities (top to bottom), for (d-f): soft microgels, (g-i): microgels of intermediate elasticity and (j-l): stiff microgels.}
\end{figure}

\begin{figure}[]
	\includegraphics[width= 3.4in]{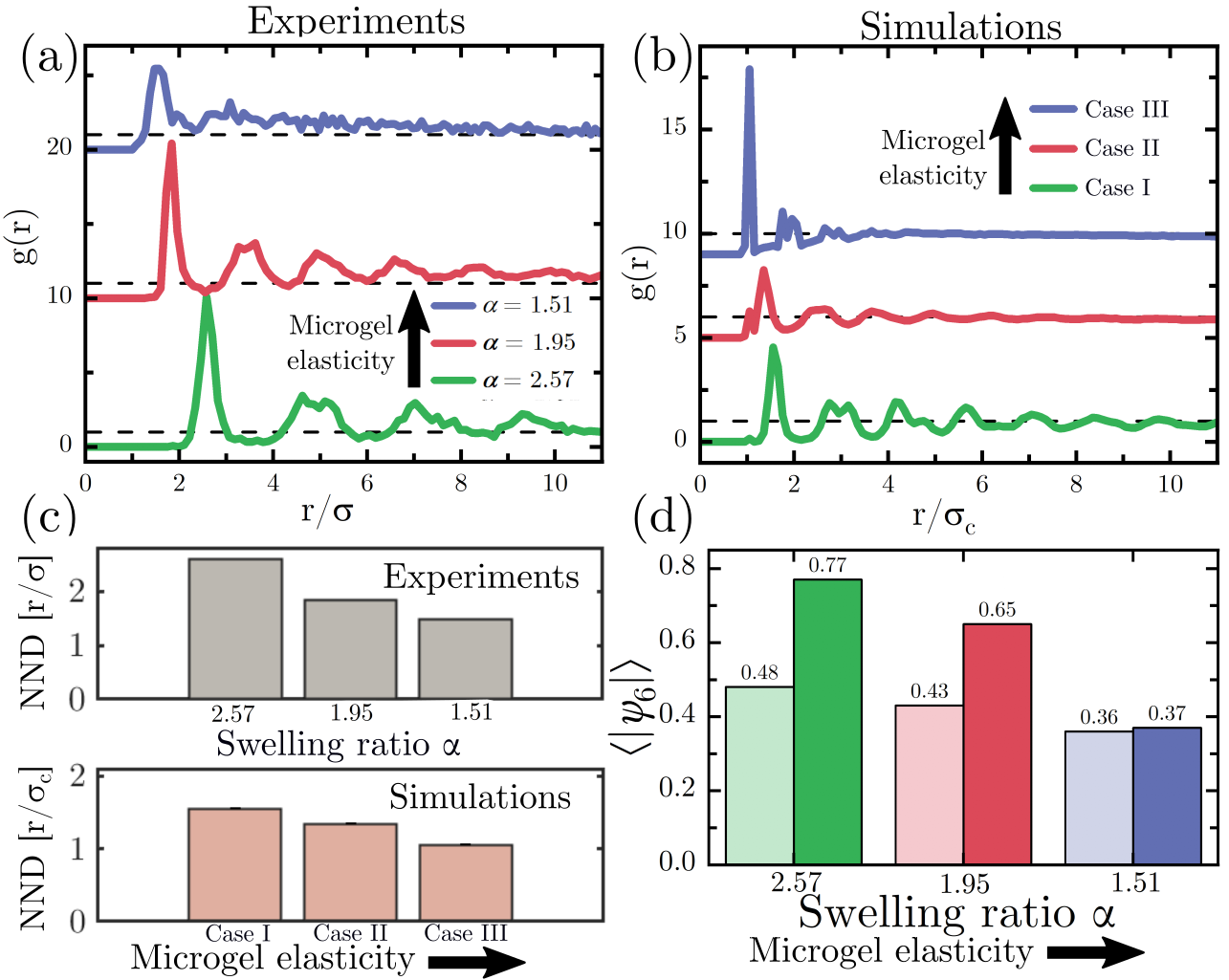}
	\centering
	\caption{\label{fig:4}  Pair correlation functions, $g(r)$, obtained from (a) microscopy images and (b) MD simulations of microgels with distinct elasticities. (c) Nearest neighbor distance (NND) as a function of microgel elasticity from experiments and simulations. (d) Evolution of $\langle |\psi_6| \rangle$ for different microgel elasticities computed for the highest $\phi$ regions in Figs.~\ref{fig:2},~\ref{fig:3}. Light and dark colored bars denote early and late times respectively. Highest $\phi$ regions (Figs.~\ref{fig:2}(f,l,r) and Figs.~\ref{fig:3}(f,i,l)) were used to calculate $g(r)$ and NND. Figs.~\ref{fig:2}(e-f,k-l,q-r) were analyzed to calculate $\langle |\psi_6| \rangle$ at early and late times.}
\end{figure}
 \vspace{-0.05cm}
Figure~\ref{fig:3}(a) displays $U(r)$ versus $r$ for soft microgels. The potential features a shallow primary minimum at a very short interparticle distance, followed by a repulsive barrier arising from steric repulsions between extended coronas. At an intermediate distance, we note a shallow (secondary) minimum due to capillary attraction. As seen from Figs.~\ref{fig:3}(d-f), this pair potential profile favors the formation of ordered clusters, voids, and non-close-packed hexagonal lattices as the microgel number density increases. With increase in microgel elasticity, the corona thickness, $d$, decreases~\cite{C5SM01743B}, while the surface charge density increases ~\cite{10.1063/5.0232833}. Consequently, enhancement in long-range dipolar repulsion between microgels of intermediate elasticity is accompanied by diminished steric repulsion. Weakening of the steric repulsion barrier between the two minima is evident from Fig.~\ref{fig:3}(b). In this pair potential, microgels can move between the secondary and primary minima, self-assembling thereby into more disordered morphologies due to the presence of two accessible length scales (Figs.~\ref{fig:3}(g-i)). Stiff microgels are characterized by shorter coronas (lowest $d$) which further reduces the interparticle steric barrier height (Fig.~\ref{fig:3}(c)). These microgels are therefore trapped irreversibly in the deep primary minimum, where they self-assemble to form anisotropic aggregates. Our MD simulations capture this behavior and replicate the transformation of chain-like structures to 2D gel-like morphologies (Figs.~\ref{fig:3}(j-l)). 

 \vspace{-0.05cm}
We computed the pair correlation functions, $g(r)$ (Fig.~\ref{fig:4}), to compare the structural features of the self-assembled microgel morphologies obtained in our experiments and simulations. The experimental curves (Fig.~\ref{fig:4}(a)) reveal a clear transition from crystalline to disordered self-organization as microgel elasticity increases. Soft microgels exhibit sharp primary and secondary peaks indicative of their long-range hexagonal order, while microgels of intermediate stiffness show weakened peaks, implying self-assembled morphologies characterized by reduced translational order. Stiff microgels display a rapid decay in $g(r)$, consistent with the observed disordered assemblies. The experimental $g(r)$ profiles are in excellent agreement with MD simulation results (Fig.~\ref{fig:4}(b)), providing robust validation for our microscopic model of inter-microgel interactions. Further inspection of Figs.~\ref{fig:4}(a,b) reveals a decrease in nearest neighbor distances as particle elasticity increases in both experiments and simulations (Fig.~\ref{fig:4}(c)). This agrees with our earlier observation that softer microgels assemble in loose, ordered structures, while stiff particles form chain-like and dense, disordered aggregates. The ordering of microgels of different elasticities at the highest $\phi$ values (Figs.~\ref{fig:2}(e-f,k-l,q-r)) was quantified by plotting $\langle |\psi_6| \rangle$ in Fig.~\ref{fig:4}(d) at early and late times (denoted respectively by light and dark color bars respectively). While soft microgel particles exhibit an increase in structural order over time, stiff microgel assemblies, kinetically trapped in chains and percolating networks, do not show any time evolution. 

 \vspace{-0.03cm}
In summary, the interface-assisted self-assembly of deformable colloids reveals a rich variety of morphologies. We show that that the spatial arrangement of microgels at the air-water interface of a drying suspension droplet is critically sensitive to both particle elasticity and local area fraction. 
As microgel elasticity increases, the experimentally observed morphologies at the interface exhibit distinct structures ranging from 2D ordered clusters, voids, and non-closed-packed hexagonal lattices to disordered 2D gel-like structures. MD simulations, implemented with a minimal effective double-well potential, successfully capture the diverse self-assembled morphologies, while revealing the key roles of elasticity-mediated steric and dipolar repulsions.

\vspace{-0.05cm}
By tuning the elasticities of core-corona systems, we have established a systematic protocol for predicting the morphologies of interfacial colloidal assemblies. In future work, it will be interesting to investigate how the competition between potential energy minima governs the collective dynamics of dense microgel phases. The presence of two length scales in the pair potential can induce frustrated dynamics even in monodisperse systems, providing a unique experimental platform to study glass physics.

\bibliography{poly}
\clearpage
\onecolumngrid %
\foreach \x in {1,...,13} {%
    \begin{center}
        \includegraphics[page=\x, width=\textwidth]{SupplementalMaterial.pdf}
    \end{center}
    \clearpage
}

\end{document}